\pdfoutput=1
\documentclass[%
 reprint,
 superscriptaddress,
 nofootinbib,
 amsmath,amssymb,
 aps,
]{revtex4-2}

\usepackage{amssymb}
\usepackage{graphicx}
\usepackage{dcolumn}
\usepackage{bm}
\usepackage[colorlinks=true,citecolor=blue,urlcolor=blue,linkcolor=red]{hyperref}
\usepackage{physics}
\usepackage{siunitx}
\usepackage{pgf}
\usepackage[normalem]{ulem}
\usepackage{xspace}
\usepackage{makecell}
\usepackage{multirow}
\usepackage{threeparttable}
\usepackage[colorlinks=true,citecolor=blue,urlcolor=blue,linkcolor=red]{hyperref}
\usepackage{orcidlink}

\newcommand{\lcdm}{\mbox{$\Lambda$CDM}}
\newcommand{\LCDM}{\mbox{$\Lambda$CDM}}

\newcommand{\mpbh}{\mbox{$M_{\rm PBH}$}}
\newcommand{\fpbh}{\mbox{$f_{\rm PBH}$}}

\newcommand{\jcap}{JCAP}

\makeatletter
\newcommand\thefontsize[1]{{#1 The current font size is: \f@size pt\par}}
\makeatother

\begin{document}

\title{Boosting the optical depth to Thomson scattering with primordial black hole evaporation at high redshift}

\author{Isaac Sierra\,\orcidlink{0000-0002-2323-303X}}
\email[]{issierra@ucdavis.edu}
\affiliation{Department of Physics and Astronomy, University of California, Davis, California, USA 95616}

\author{Gabriel P. Lynch\,\orcidlink{0009-0004-3143-1708}}
\affiliation{Department of Physics and Astronomy, University of California, Davis, California, USA 95616}

\author{Lloyd Knox}
\affiliation{Department of Physics and Astronomy, University of California, Davis, California, USA 95616}

\author{Vivian Poulin\,\orcidlink{0000-0002-9117-5257}}
\affiliation{Laboratoire Univers et Particules de Montpellier (LUPM),
CC 72, Place Eug\`ene Bataillon, 34095 Montpellier - Cedex 5, France}

\author{Lennart Balkenhol\,\orcidlink{0000-0001-6899-1873}}
\affiliation{Sorbonne Universit\`e, CNRS, UMR 7095, Institut d’Astrophysique de Paris, 98 bis bd Arago, 75014 Paris, France}

\author{Ali Rida Khalife\,\orcidlink{0000-0002-8388-4950}}
\affiliation{Sorbonne Universit\`e, CNRS, UMR 7095, Institut d’Astrophysique de Paris, 98 bis bd Arago, 75014 Paris, France}

\date{\today}

\begin{abstract}

BAO and CMB data are somewhat discrepant when interpreted in the context of \lcdm, discrepancies that show up as a `matter density deficit' and as a `CMB lensing excess'. One possible resolution is an increased optical depth to scattering off of free electrons in the post-recombination universe, $\tau$, a possibility raised by Sailer et al. 2025 and Jhaveri et al. 2025. Since Planck measurements of the low-$\ell$ polarization `reionization bump' already constrain $\tau$ from standard stellar-driven reionization at $z<10$, we investigate additional optical depth sourced by transient or partial reionization at higher redshift from exotic processes. For specificity, we explore the impact of Hawking radiation from a monochromatic spectrum of primordial black holes, retaining the high-$\ell$ $TT/TE/EE$ data that constrain such histories and varying the reionization redshift jointly. We find that the CMB data do not significantly prefer these additional signals: the boost is at most $\Delta\tau \simeq 0.008$, well short of the $\Delta\tau \simeq 0.03$ that would completely eliminate the moderate discrepancy. The matter density deficit and the lensing excess are not significantly eased: we explain why, tracing it to compensation from the reionization redshift and the residual PBH signal at $\ell > 30$.
\end{abstract}

\maketitle

\section{\label{sec:intro}Introduction}

Baryon acoustic oscillation (BAO) data from DESI DR2 \cite{DESI:2025zgx} and CMB data from Planck, ACT, and SPT \cite{Planck:2019nip, carron_cmb_2022, ACT:2023dou, 2025arXiv250314452L, SPT-3G:2024atg, SPT-3G:2025bzu} are in mild tension with each other when interpreted using the \lcdm\ model. The discrepancy manifests in two related ways: a \emph{matter density deficit}, in which the present-day non-relativistic matter density inferred from the CMB exceeds that inferred from BAO data\footnote{The matter density deficit is most significant when the angular size of the sound horizon at the end of the baryon drag epoch, inferred from CMB data, is added to the BAO data \cite{Loverde:2024nfi}.}, and a \emph{lensing excess}, in which the lensing power measured both from lensing reconstruction and from its smoothing of the temperature and polarization two-point functions exceeds the prediction of the combined CMB$+$BAO fit \cite{Loverde:2024nfi,Lynch:2025ine}\footnote{See also \cite{Weiner:2026sfm} for a compact framing of the tensions as a discrepancy in the ``matter era distance indicator."}. The same tension complicates the interpretation of neutrino-mass constraints \cite{Craig:2024tky,Green:2024xbb,Loverde:2024nfi,Lynch:2025ine} and, if BAO and CMB data are combined under \lcdm\ regardless, carries implications for inflationary model building and the search for primordial gravitational waves \citep[e.g.][]{Ferreira:2025lrd}. A number of extensions reduce the tension, including dynamical dark energy \citep{Elbers:2024sha,2025Rodrigues_DDE,2026Wang_DDE}, interacting dark matter and dark energy \citep{Teixeira:2024qmw, Khoury:2025txd}, decaying dark matter \citep{Lynch:2025ine, Montandon:2026vuc}, early recombination (perhaps driven by primordial magnetic fields)\citep{Lynch:2024hzh, Mirpoorian:2025rfp}, and early dark energy \cite{Poulin:2025nfb,SPT-3G:2025vyw,Garny:2025kqj}.

\citet{Sailer:2025lxj} and \citet{Jhaveri:2025neg} pointed out a less exotic possibility:
if the $\ell < 30$ CMB polarization data are excluded, the BAO--CMB tension is greatly reduced
and the inferred optical depth to Thomson scattering rises to $\tau \simeq 0.09$. The difficulty
is that such a large $\tau$, if produced by reionization at $z \lesssim 10$, would enhance the
`reionization bump' in the large-scale polarization power spectrum at $\ell \sim 5$--$10$ well
beyond what is observed; the low-$\ell$ data accordingly disfavor $\tau \simeq 0.09$ at high
significance, yielding $\tau = 0.051 \pm 0.006$ \cite{Planck:2020olo}. This result, from the NPIPE (PR4) large-scale polarization, is the lowest of the modern Planck determinations; reanalyses of the same data span $\tau \simeq 0.051$--$0.063$, depending on the map processing and on whether high-$\ell$ temperature and polarization information is included \cite{Pagano:2019tci,deBelsunce:2021mec,Rosenberg:2022sdy,Tristram:2023haj,Genesini:2026lmg}. Note that \citet{Genesini:2026lmg} provide a helpful summary of many of these analyses, as well as valuable robustness tests with their combined use of WMAP, LFI, and HFI data. The ground-based experiment, CLASS, has also produced a constraint on $\tau$ by crossing their maps \cite{Eimer:2023esh} with Planck to find $\tau = 0.053^{+0.018}_{-0.019}$ \cite{CLASS:2025khf}.

A high $\tau$ is likewise in apparent conflict with spectroscopic inferences from high-redshift
quasars and galaxies. Measurements of the neutral-hydrogen fraction from the Gunn--Peterson trough
and the Lyman-$\alpha$ damping wing point to a late and rapid reionization, and hence to
$\tau \simeq 0.05$. Using a compilation of such data\footnote{Together with BAO and BBN data to fix the mean
hydrogen density and $H(z)$ entering the optical-depth integral.} \citet{Elbers:2025xvk} finds
$\tau = 0.0492^{+0.0014}_{-0.0030}$.
Consistent with this, \citet{Hazra:2019wdn} 
found that replacing the Planck low-$\ell$ $EE$ likelihood with astrophysical
tracers of the reionization history recovers a mean optical depth in agreement
with the CMB determination.

These constraints, however, bear principally on the \emph{low-redshift} contribution to $\tau$.
The optical depth is an integral over the entire post-recombination free-electron history, and
different observables weight that history by redshift in different ways. The low-$\ell$ bump and the spectroscopic data both constrain the $z \lesssim 10$ contribution, though the latter are themselves limited by the absence of quasar data at higher redshift. Free electrons at higher redshift still add to $\tau$, but the polarization they generate appears at correspondingly higher multipoles. A quadrupole scattering off free electrons at redshift $z$ predominantly sources $E$-mode anisotropies on the angular scale of the horizon at that $z$, projected to the present, so that raising $z$ moves the bulk of the induced signal to higher $\ell$. A high-redshift ionization component can therefore boost $\tau$ toward $0.09$ while leaving both the $\ell \lesssim 10$ bump and the spectroscopic picture largely intact---but it is not unconstrained: it imprints on $EE$ and $TE$ at $\ell > 30$ and well beyond, and, because the $\exp(-2\tau)$ suppression of the primary
anisotropies is complete only for Fourier modes below the horizon scale of the scattering, that suppression remains incomplete through $\ell \sim 30$ for high-redshift electrons---an imprint carried in $TT$ as well.

The possibility of a high-redshift contribution to $\tau$ has a long history in CMB analysis.
\citet{Holder:2003eb} first noted that $\tau$ estimated from low-$\ell$ polarization can be biased
if an extended, high-redshift tail of reionization is neglected. The principal-component framework
of \citet{Mortonson:2007hq}, building on \citet{Holder:2003eb}, and the CMBPol forecasts of \citet{Zaldarriaga:2008ap} 
formalized how the large-scale and $\ell \sim 10$--$20$ $E$-mode data separate the $z<10$ and $z>10$
contributions to $\tau$. Reconstructions of the Planck data subsequently debated whether they prefer
a $z \gtrsim 15$ contribution \cite{Heinrich:2016ojb, Heinrich:2018btc} 
or are consistent with none \cite{Millea:2018bko, Planck:2020olo},
with the free-form (``FlexKnot'') analysis of \citet{Millea:2018bko} placing upper limits on the partial
optical depth $\tau(15 < z < 30) \lesssim 0.015$. Combining the CMB with astrophysical
reionization tracers rather than the low-$\ell$ $EE$ likelihood, \citet{Paoletti:2021gzr} bound the
same quantity to $\tau(15 < z < 30) < 0.001$ in their most flexible, non-monotonic reconstruction.
These analyses, however, generally truncated at $z \sim 30$ and predate the BAO--CMB tension.

Since \citet{Sailer:2025lxj} and \citet{Jhaveri:2025neg}, several works have revived high-redshift
ionization specifically as a means of boosting $\tau$ to relieve the tension. \citet{Tan:2025obi} 
propose a transient ``flash'' of Population~III.1 stars creating ionization at $z \sim 20$--$25$; \citet{Cheng:2025cmb}
reconstruct the history to high redshift with Gaussian processes; \citet{Ilic:2025idl} fit a ten-bin
model over $5 < z < 25$; and \citet{Yin:2026hvw} 
treat PBH Hawking evaporation as a high-redshift ionization source. Each, however, leaves open the
regime we target: the Population~III.1 studies do not confront the polarization data at the likelihood
level; the Gaussian-process reconstruction (and therefore \citet{Yin:2026hvw} which starts from that reconstruction) 
uses only the low-$\ell$ ($\ell \le 30$)
$E$-mode likelihood and so cannot capture the higher-$\ell$ imprint of the high-redshift electrons; and
\citet{Ilic:2025idl} cap their reconstruction at $z = 25$. Additionally, we note that \citet{Yin:2026hvw} are
focused on constraining PBH scenarios, and so do not perform an estimate of cosmological parameters to assess
any possible resolution of the BAO--CMB tension in their model space.

The free electrons required at high redshift could be freed from atoms by radiation from sources such as evaporating primordial
black holes (PBHs) \cite{Poulin:2016anj,Acharya:2020jbv,Auffinger:2022khh,Carr:2020gox}, accretion onto
stellar-mass PBHs \cite[e.g.][]{Poulin:2017bwe}, decaying or annihilating particles
\cite{Chen:2003gz,Padmanabhan:2005es}, or Population~III.1 stars \cite{Tan:2025obi}. The absence of the
associated polarization signal has long been used to \emph{bound} such energy injection
\cite{Galli:2009zc,Slatyer:2009yq,Planck:2015fie,Slatyer:2016qyl,Planck:2018vyg}. Here we ask a related yet different question: how
large a boost to $\tau$ could such a scenario produce while remaining consistent with the data? For specificity in answering this question, we consider ionizations due to the Hawking radiation from a monochromatic population of PBHs.

We study the model space of \lcdm\ augmented by a single PBH mass, \mpbh, and the PBH fraction of the dark matter, $f_{\rm PBH}$; black holes of the masses we consider inject energy over a range extending up to recombination. Our primary analysis uses Planck PR3, and we consider additional combinations of PR3 with the Planck, ACT, and SPT lensing reconstructions and the ACT~DR6 and SPT-3G $TT/TE/EE$ spectra, with and without BAO. 

Our treatment confronts a high-redshift reionization scenario with high-multipole data: we forward-model a physical injection scenario whose free electrons extend beyond $z = 30$, all the way to recombination, while retaining the high-$\ell$ $TT/TE/EE$ data that constrain such a scenario. Because we forward-model a physical history rather than reconstructing
$X_{\rm e}(z)$ in a truncated principal-component basis, we also do not discard the poorly-constrained high-redshift modes that a PCA truncation removes \cite{Mortonson:2007hq}---precisely the modes capable of raising $\tau$. Finally, retaining the high-$\ell$ data lets us go beyond the shift in $\tau$ to check the bottom line: whether the CMB lensing excess and matter density deficit are alleviated.

We begin in Section~\ref{sec:preliminaries} by briefly summarizing the PBH model that we use and presenting some illustrative ionization histories and their corresponding CMB spectra, before discussing the data sets we use in our analysis (Section~\ref{sec:data}). In Section~\ref{sec:results} we present our results on raising the optical depth with evaporating PBHs, before discussing and concluding in Section~\ref{sec:conclusions}.

\section{Preliminaries}\label{sec:preliminaries}

\subsection{Primordial black hole model}

For all of our theoretical computations we utilize the PBH parameterization of \citet{Poulin:2016anj,Stocker:2018exoclass}, publicly available in \texttt{CLASS} \citep{2011JCAP...07..034B}, and we restrict our analysis to low mass PBHs ($10^{13}$-$10^{15}$ grams). In this regime, the PBH lifetime places the bulk of evaporation after recombination, and accretion onto the PBHs is negligible, so that Hawking evaporation dominates the energy injection. We make several further simplifying assumptions, whose consequences we discuss below. We assume: (i) that the PBH emission is entirely into photon and $e^{\pm}$ channels; (ii)
that injected energy is deposited ``on-the-spot" (see \citet{Poulin:2016anj,Stocker:2018exoclass}); (iii) that the energy injection is governed by the standard Hawking evaporation rate for non-rotating (Schwarzschild) black holes; and finally, (iv) that the PBHs have a monochromatic mass distribution.

The Schwarzschild emission spectrum is set by the black hole's surface area and Hawking temperature ($T \propto M^{-1}$) and results in the mass-loss scaling $\dv*{M}{t} \propto M^{-2}$ \citep{Hawking:1974rv, Hawking:1975vcx}. Therefore, $M$ controls when the evaporation, and hence the energy injection, takes place. The injection rate in this case is \citep{Poulin:2016anj}:

\begin{equation}
    \frac{\dd E}{\dd V \dd t}\bigg|_{\rm inj,PBH} = \frac{\Omega_{\rm cdm}\rho_{c}(1+z)^{3}f_{\rm PBH}}{M_{\rm PBH}^{\rm ini}} \eval{\dv{M}{t}}_{\rm e.m.},
    \label{eq:pbh_injection}
\end{equation}
where $\eval{\dv*{M}{t}}_{\rm e.m.} = f_{\rm e.m.} \dv{M}{t}$ is the mass-loss rate weighted by the electromagnetic branching ratio $f_{\rm e.m.}$. We compute the mass-loss rate accounting for all channels open at a given $T$, as in \cite{Poulin:2016anj,Stocker:2018exoclass}. However we make the simplifying assumption that 
$f_{\rm e.m.} =1$. This overestimates the injection at the 10-50\% level in the mass range we consider, but can be captured by a trivial rescaling of $f_{\rm PBH}$.

Indeed, the data are only sensitive to these parameters insofar as they source a given ionization history $X_{\rm e}(z)$. That history depends not on the injection rate (Equation~\ref{eq:pbh_injection}) but on the energy {\it deposition} rate, which is related to Equation~\ref{eq:pbh_injection} by the deposition efficiency $f_c(z, X_{\rm e})$ \citep{Finkbeiner:2011dx, Slatyer:2012yq, Poulin:2016anj}. The 
energy deposition term modifying $X_{\rm e}(z)$ can therefore be written 
as

\begin{eqnarray}
    \frac{\dd E}{\dd V \dd t}\bigg|_{\rm dep,PBH}& \equiv & \frac{\dd E}{\dd V \dd t}\bigg|_{\rm inj,PBH}\times f_c(z,X_e)\\ 
    & \propto &  \left(f_{\rm PBH} f_{\rm e.m.} \right) \times \left( \frac{1}{M_{\rm PBH}^{\rm ini}} \dv{M}{t}(z; M_{\rm pbh})\right) \times f_c(z, X_{\rm e}).\nonumber
    \label{eqn:xe_source}
\end{eqnarray}

The PBH abundance $f_{\rm PBH}$ and $f_{\rm e.m.}$ only enter via their product as an overall amplitude, and furthermore $f_c(z, X_{\rm e})$ enters as a third amplitude factor that is weakly dependent on $X_{\rm e}$. Changes to $f_{\rm e.m.}$ are therefore completely degenerate with a rescaling of $f_{\rm PBH}$, in which case the achievable $X_{\rm e}(z)$ (and hence $\tau$) is unchanged. For $f_c$ this degeneracy is not exact, since the $X_{\rm e}(z)$ dependence could in principle alter the shape of $X_{\rm e}$; in Appendix~\ref{sec:BeyondOn-the-spot} we verify this is not the case  by comparing to a ``beyond on-the-spot" deposition treatment. Therefore, while our assumptions (i) and (ii) impact our inferred limits on $f_{\rm PBH}$, these limits are not our primary interest: for the quantities that {\it are} of interest (e.g. $X_{\rm e}, \tau$), changing either of these assumptions simply shifts the relevant regime for $f_{\rm PBH}$.

Our two additional assumptions are not captured by this argument. An extended mass function amalgamates evaporation histories with different peak redshifts and can impact the shape of the resulting $X_{\rm e}(z)$ rather than only its amplitude. Similarly, Kerr black holes have a different emission spectrum and lifetime than the Schwarzschild case. We leave a quantitative treatment of these more complex cases for future work.

Finally, we note that such post-recombination electromagnetic energy injections generically lead to CMB spectral distortions, in particular of the $y$-type (e.g. \cite{Chluba:2011hw,Chluba:2013wsa}). However, these stay orders of magnitude below the FIRAS bound $y<1.5\times10^{-5}$ \cite{Fixsen:1996nj,2009ApJ...707..916F} for the $f_{\rm PBH}$ and \mpbh~values considered in this work \cite{Poulin:2016anj,Stocker:2018exoclass,Acharya:2020jbv}.

\subsection{Reionization history}

\begin{figure*} 
    \centering
    \includegraphics[scale=0.38]{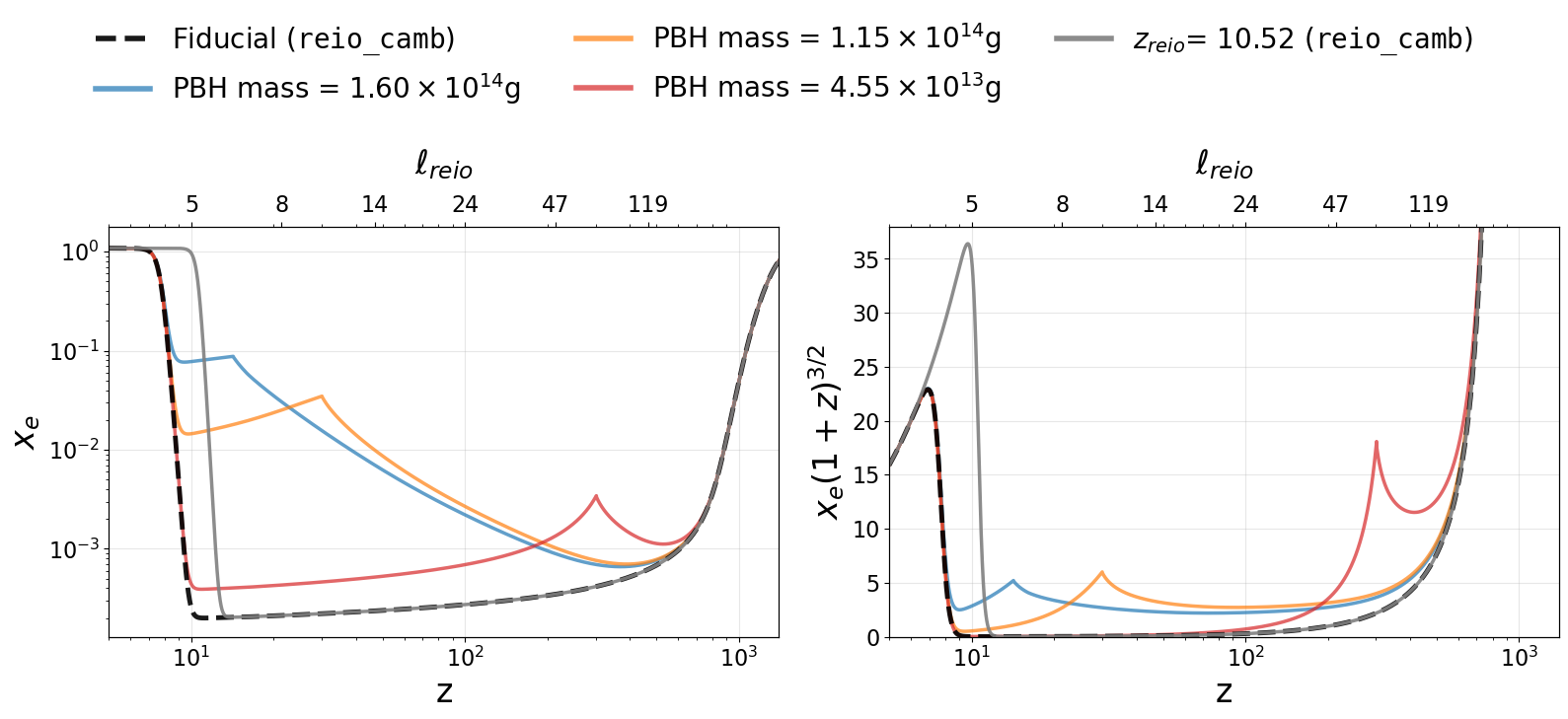}
    \hspace{-2mm}
    \centering
    \includegraphics[scale=0.48]{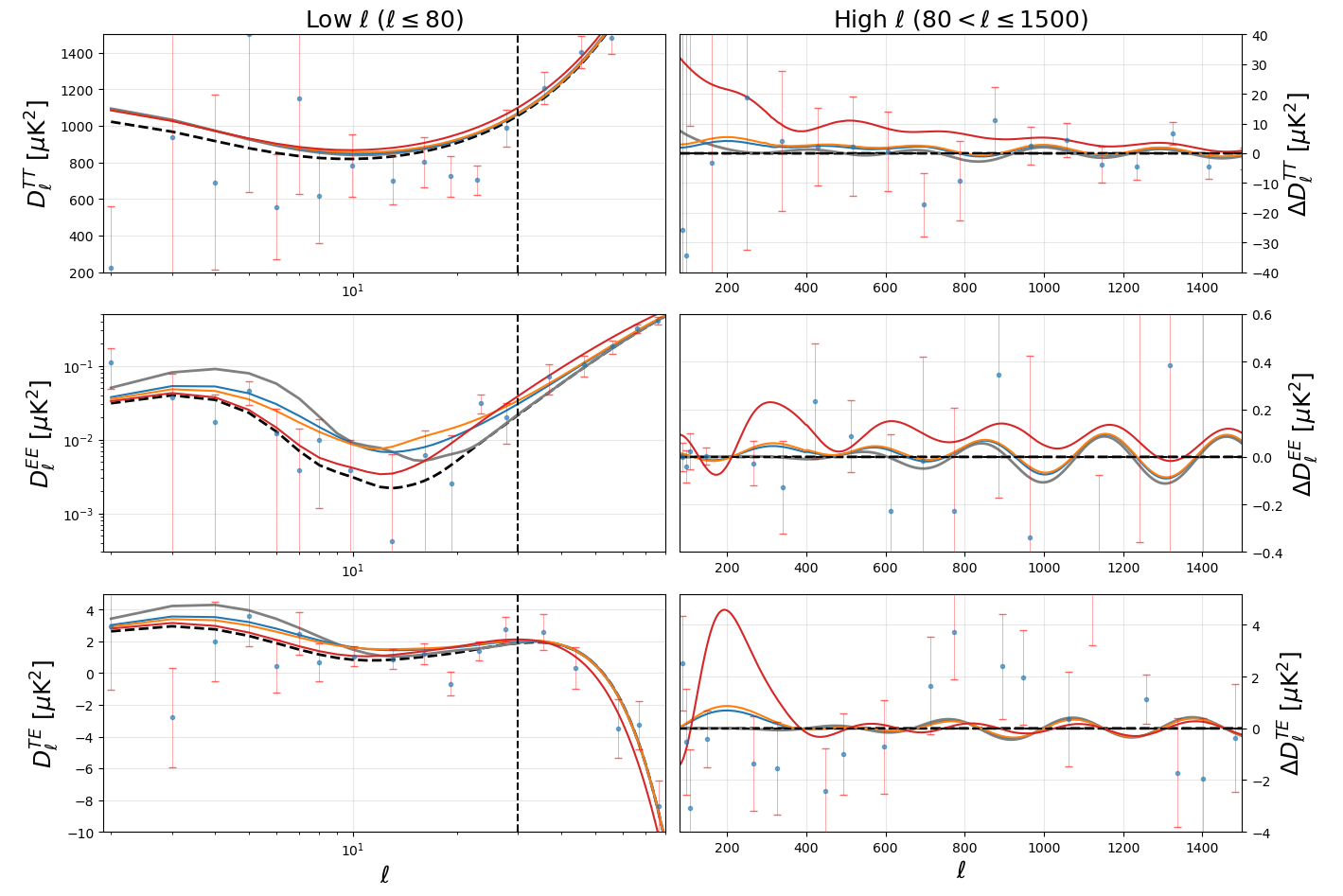}        
    \caption{Ionization histories for a set of PBH masses with peak evaporations at $z=$ 15, 30, and 300, with PBH fraction adjusted for each to yield an increase in $\tau$ of 0.03. Two $\Lambda$CDM cases are also shown: the fiducial model and one with the tanh step shifted to higher $z$ to yield an increase in $\tau$ of 0.03 as well. We also indicate the approximate peak multipole of the $EE$ reionization bump, $\ell_{reio}$ for the corresponding redshifts of peak $X_{\rm e}(z)$ (see text). We show corresponding angular power spectra in the six lower panels, together with binned Planck power spectrum measurements. The right panels show these at higher $\ell$ as residuals  relative to the fiducial model. All of these model spectra are for models with $H_0$, $\omega_b$, $\omega_{\rm c}$, $n_s$, and $A_s \exp(-2\tau)$ fixed to their best-fit Planck \lcdm\ values.}
    \label{fig:Spectra}
\end{figure*}

In the top two panels of Fig.~\ref{fig:Spectra}, we show ionization histories from a sample of black hole masses, chosen to span $10 \lesssim z_{\rm peak} \lesssim 300$, where $z_{\rm peak}$ is the redshift of peak increase to $X_{\rm e}(z)$. For comparisons we include two reionization histories with approximate step functions. These are for our fiducial model, \lcdm \ using the Planck 2018 baseline mean cosmological parameter values: \{$\omega_{\rm b}$: $0.02237$, $\omega_{\mathrm{c}}$: $0.1200$, $n_s$: $0.9649$, $A_s$: $2.1 \times 10^{-9}$, $h$: $0.6736$, $z_{\mathrm{reio}}$: $7.67$\}, and for one with a higher redshift of reionization, $z_{\rm reio} = 10.52$. All of the non-fiducial models here increase $\tau$ by 0.03 above the $\tau = 0.055$ of the fiducial model. 

In addition to plotting $X_{\rm e}(z)$ vs. $z$ with a logarithmic $y$ axis, we also show $X_{\rm e}(z) \times (1+z)^{3/2}$ vs. $z$ with a linear $y$ axis. The latter is useful since constant $X_{\rm e}(z) \times (1+z)^{3/2}$ gives equal contributions to $\tau$ from any logarithmic interval in $z$, throughout the matter-dominated era. We note that the $(1+z)^{3/2}$ factor increases the relative importance of $z > z_{\rm peak}$ so that for models with higher mass, and hence later evaporation and lower $X_{\rm e}(z_{\rm peak})$, the contributions to $\tau$ come from a very broad range in redshifts, extending back to the tail end of the recombination era.  Lower-mass models, which evaporate earlier, have their overall emission history more concentrated at higher redshift with a proportionally larger share of their $\tau$ contribution arising from this post-recombination tail.

\subsection{Impact on the primary CMB}
\label{sec:ImpactOnCMB}

In the six lower panels of Fig.~\ref{fig:Spectra} we see the corresponding power spectra. All models are for fixed $\omega_b, \omega_c, H_0$, and $n_s$. We also fix $A_s \exp(-2\tau)$ anticipating that a fit to the data would exploit this quasi-degeneracy between $A_s$ and $\tau$. 

Let us first consider the spectra for the $z_{\rm reio} = 10.52$ model. As expected, the power at low $\ell$ in $EE$ and $TE$ is increased relative to the fiducial model. We also note that all the visible differences at $\ell \gtrsim 200$, with the fiducial model spectra, are almost entirely due to the increased gravitational lensing power of the $z_{\rm reio} = 10.52$ model, due to the larger value of $A_s$. These differences are visible here starting at $\ell \simeq 650$ in both $TE$ and in $EE$, and somewhat in $TT$ as well. 

The upward adjustment in $A_s$ boosts all the non-fiducial spectra relative to the fiducial spectra on angular scales larger than the horizon at the earliest significant density of free electrons from reionization. For the tanh model this is at $\ell \lesssim 20$. This is clearest in $TT$ where the only effect of the additional electron scattering is suppression. For ionization histories with preferentially more scattering from high redshifts, the transition to the asymptotically valid $\exp(-2\tau)$ value for the suppression (which cancels out the boost we applied to $A_s$) is at smaller angular scales. Note that even the PBH models with $X_e(z)$ peaks at redshifts as low as 15 and 30 create significant optical depth at redshifts all the way back to the tail end of the recombination epoch, as can be seen in the top right panel, which explains the very slow approach, as $\ell$ increases, toward the asymptotically expected $\exp(-2\tau)$ suppression.

We also see in the $EE$ and $TE$ PBH spectra the expected reduction, relative to the $z_{\rm reio} = 10.52$ model, of power at $\ell \lesssim 10$ for our highest-mass model and at $\ell \lesssim 20$ for our lowest-mass model. This expected reduction in power is our motivation for exploring higher-redshift reionization as a possible means of increasing $\tau$. 

Polarization generation in these models shows up at higher $\ell$, where we see increases in power instead. The multipole moment of peak polarization power from free electrons at redshift $z$ has been approximated as \citep{Knox:2003ch}
\begin{equation} 
\ell_{reio} \approx 2\frac{\eta_0 - \eta(z)}{\eta(z) - \eta(z_*)},
\label{eqn:ellreio}
\end{equation} 
where $\eta$ is conformal time, $z_*$ is the redshift of recombination, and we have assumed zero mean curvature. This $\ell_{\rm reio}$ is included on the top horizontal scale of the top two panels.
The higher $\ell$ imprint occurs for two reasons: 1) the temperature quadrupole seen by higher-redshift electrons is generated by shorter-wavelength modes, an effect that shows up in the denominator of the above expression, and 2) these modes project into angular scales over a larger distance, given in the numerator. We see the two tanh models producing reionization bumps that fit this prescription, but for the PBH models the reionization signals are more complicated and $\ell_{\rm reio}$ offers only a rough guide to their nature. Note that effects of the additional free electrons show up both at $\ell \lesssim 30$ and at multipole moments well beyond 30 (and $\ell_{\rm reio}$). Only at high multipoles do we see the PBH $\Delta \tau = 0.03$ model spectra approach the $z_{\rm reio} = 10.52$ spectra. 

These changes in power at $\ell\, >\, 30$, beyond what we get from the increase in lensing power, mean that the high-$\ell$ data are important to include, in an investigation of a $\tau$-focused solution to the BAO--CMB tension, for two reasons.  First, these signals can potentially provide further constraints on these scenarios, affecting how much we can exploit them to increase $\tau$. Second, because of these high-$\ell$ signals, the impact of these scenarios on cosmological parameters does not flow solely through the $\tau - A_s - \omega_{\rm c}$ degeneracy direction (see Section~\ref{sec:Shifts} for details). The high-$\ell$ signals will cause additional shifts in cosmological parameters, which can only be inferred by inclusion of the high-$\ell$ data in our analyses. 

\subsection{Data, likelihoods, and priors}
\label{sec:data}

We perform Bayesian parameter inference using Markov Chain Monte Carlo (MCMC)
methods \citep{Christensen:2001gj} within the \texttt{Cobaya} framework
\citep{Torrado:2020dgo}, and analyze the resulting chains with \texttt{GetDist}
\citep{Lewis:2019xzd}.

Our baseline dataset is the Planck 2018 (PR3) CMB power spectra alone. Among
current CMB experiments, Planck is the one indispensable dataset for this study:
its near cosmic-variance-limited measurement of the large-scale $E$-mode
polarization is what pins the reionization optical depth $\tau$, and the shape of
the large- and intermediate-scale $E$-mode signal is what carries information on
the \emph{redshift distribution} of the free electrons that source it. We
therefore use Planck alone to establish our constraints on the ionization history
and on $\tau$. The baseline comprises the high-$\ell$
($30 \le \ell \lesssim 2500$) \texttt{plik\_lite} $TT/TE/EE$ likelihood, with
foreground nuisance parameters pre-marginalized \citep{Planck:2019nip}; the
low-$\ell$ ($2 \le \ell < 30$) \texttt{Commander} temperature likelihood; and the
low-$\ell$ \texttt{SimAll} $E$-mode likelihood \citep{Planck:2019nip}.

The low-$\ell$ $E$-mode likelihood is the most consequential choice for our
analysis, since it dominates the constraint on $\tau$. \texttt{SimAll} is a
simulation-based likelihood constructed from the Planck 100 and 143\,GHz $E$-mode
cross-spectra; it provides $EE$ (and $BB$) but not $TE$. The low-$\ell$ $TE$
spectrum is excluded from the Planck baseline because it carries little
statistical weight relative to $EE$ on these scales and exhibits excess scatter of
not-fully-understood origin at a few low multipoles \citep{Planck:2019nip}.
Alternative low-$\ell$ $E$-mode likelihoods are available, including the
HFI-reprocessed \texttt{SRoll2} likelihood \citep{Pagano:2019tci} and the
PR4/NPIPE-based \texttt{LoLLiPoP} likelihood \citep{Tristram:2023haj}, which
recover mean values of $\tau$ differing at the few$\,\times 10^{-3}$ level. To test
the sensitivity of our conclusions to this choice, we repeat our baseline analysis
with \texttt{SRoll2} substituted for \texttt{SimAll} \citep{Pagano:2019tci}.

A high-redshift contribution to $\tau$ does more than shift $\tau$ itself: by
rescaling the amplitude of the primary anisotropies and modifying the lensing
power, it shifts the inferred \lcdm\ parameters. These shifts are influenced by
CMB lensing data, and by smaller-scale power spectrum measurements. These parameter
shifts are also highly relevant to our motivation for this investigation: addressing
the BAO--CMB tension. For further investigation of these shifts we therefore also conduct analyses that
add ground-based small-scale power spectra, CMB lensing, and BAO data. For the temperature
and polarization spectra we add ACT DR6 \citep{2025arXiv250314452L} and SPT-3G D1
\citep{SPT-3G:2025bzu}, the latter via the \texttt{candl} likelihood framework \citep{Balkenhol:2024candl}. Where ACT DR6 is combined with Planck we follow the ACT
Collaboration's ``P-ACT'' prescription, in which the Planck PR3 high-$\ell$
\texttt{plik\_lite} spectra are retained only at $\ell < 1000$, $600$, and $600$
in $TT$, $TE$, and $EE$ respectively, with ACT DR6 supplying the smaller-scale
information; this avoids double-counting the modes measured in common by the two
experiments \citep{2025arXiv250314452L}. We retain our baseline low-$\ell$
likelihoods (\texttt{Commander} and \texttt{SimAll}) throughout. For CMB lensing we
use the combined Planck PR4 and ACT DR6 lensing reconstruction \citep{carron_cmb_2022,ACT:2023kun,ACT:2023dou},
both on its own and together with the SPT-3G polarization lensing reconstruction
\citep{SPT-3G:2024atg}. The combination of all three of these lensing reconstructions was explored in \citet{SPT-3G:2025zuh} where it was called ``APS Lensing." For BAO we use the DESI DR2 BAO measurements \citep{DESI:2025zgx}.

We refer to the datasets we use as:
\begin{itemize}
  \item Planck: Planck PR3 \texttt{plik\_lite} high-$\ell$ $TT/TE/EE$ $+$
        \texttt{Commander} low-$\ell$ $TT$ $+$ \texttt{SimAll} low-$\ell$ $EE$.
        This is our baseline dataset.
  \item Planck (SRoll2): the baseline with \texttt{SRoll2} in place of
        \texttt{SimAll}.
  \item SPA: Planck augmented with ACT DR6 and SPT-3G D1
        $TT/TE/EE$; the Planck$+$ACT combination follows the P-ACT prescription
        (Planck \texttt{plik\_lite} retained only at $\ell < 1000/600/600$ in
        $TT/TE/EE$), with the baseline low-$\ell$ likelihoods retained.
  \item SPA Lensing: Lensing reconstruction from Planck PR4, ACT DR6 and the SPT-3G polarization.
    \item DESI: DESI DR2 BAO.
    \item Acoustic Scale: DESI + $\theta_{\rm s}$ as inferred from SPA under the assumption of \lcdm. 
\end{itemize}

This final dataset, Acoustic Scale, is an extension of our BAO data to include an inference of $\theta_{\rm s}$ -- the angular size of the sound horizon at the end of the baryon drag epoch. We use a result for this dataset from \citet{Weiner:2026sfm} where $\theta_s$ is inferred from SPA under the assumption of \lcdm. 

One consequence of our baseline analysis only using Planck data, and only Planck
CMB power spectra, is that the
BAO--CMB tension has a significance smaller than $2\sigma$, which one might 
reasonably call no tension at all. So it can appear we are addressing a 
non-existent tension. However, the extension to SPA plus SPA Lensing 
increases the significance of the tension to nearly $3 \sigma$
\cite{SPT-3G:2025bzu}.

For our Bayesian analyses we need to adopt prior distributions for all our model parameters. Especially important are those for the PBH parameters. For these we adopted the priors shown in Table~\ref{tab:pbh_priors}. 

Our uniform priors on $f_{\rm PBH}$ are motivated by the CMB limits of \citet{Carr2010}. We cap the prior at their bound evaluated at the upper edge of our mass range, $M_{\rm PBH} = 3\times10^{14}\,{\rm g}$, and set the lower bound to $1\times10^{-12}$, corresponding to the regime of negligible PBH impact. The PBH mass prior is bounded at the upper end by masses corresponding to incomplete evaporation; this produces a gradual transition into the $\tanh$ step, where $M_{\rm PBH}$ becomes strongly degenerate with $f_{\rm PBH}$ (see Section{~\ref{sec:MatterAndLensing}). The lower mass bound corresponds to an evaporation redshift of $z\sim300$; exploratory runs testing evaporation as early as $z\sim500$ produced no additional phenomenological insight, justifying this cutoff. 

\begin{table}[h]
\centering
\begin{threeparttable}
\setlength{\tabcolsep}{1.3pt}
\renewcommand{\arraystretch}{1.2}
\scriptsize
\small
\begin{tabular}{lcc}
\hline
Model & $M_{\rm PBH}$ [$10^{14}$ g] & $f_{\rm PBH}$ \\
\hline
$\Lambda$CDM + $f_{\rm PBH}$ + \mpbh & $0.45 - 3.0$ & $10^{-12} - 5\times10^{-7}$ \\
$\Lambda$CDM + $f_{\rm PBH}$ + $A_{\rm lens}$ & --          & $10^{-12} - 5\times10^{-7}$ \\
\hline
\end{tabular}
\end{threeparttable}
\caption{Priors adopted for the PBH parameters in each model space.}
\label{tab:pbh_priors}
\end{table}

\begin{figure*} 
    \centering
    \includegraphics[scale=0.45]{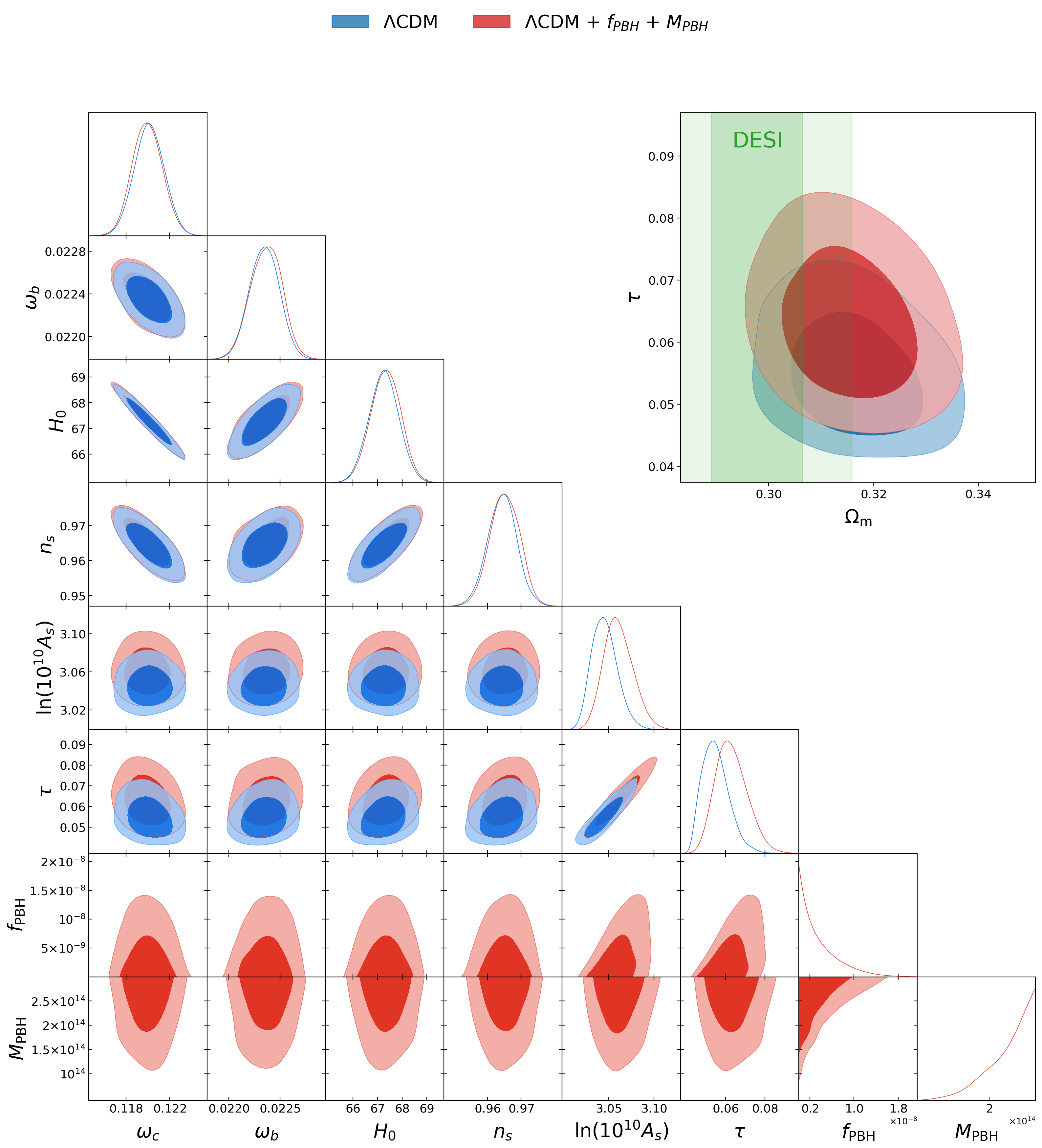}
    \caption{\textbf{Bottom Left:} Cornerplot for $\Lambda$CDM + PBH (with both evaporation mass and fraction varying with uniform priors) for the Planck likelihood. \textbf{Top Right:} Marginal 2D posterior in the $\Omega_m$ - $\tau$ plane for the fiducial $\Lambda$CDM model and the $\Lambda$CDM + PBH model, with DESI DR2 $\Omega_m$ posterior plotted for reference. }
    \label{fig:Cornerplot}
\end{figure*}

\begin{figure} 
    \centering
    \includegraphics[width=\columnwidth]{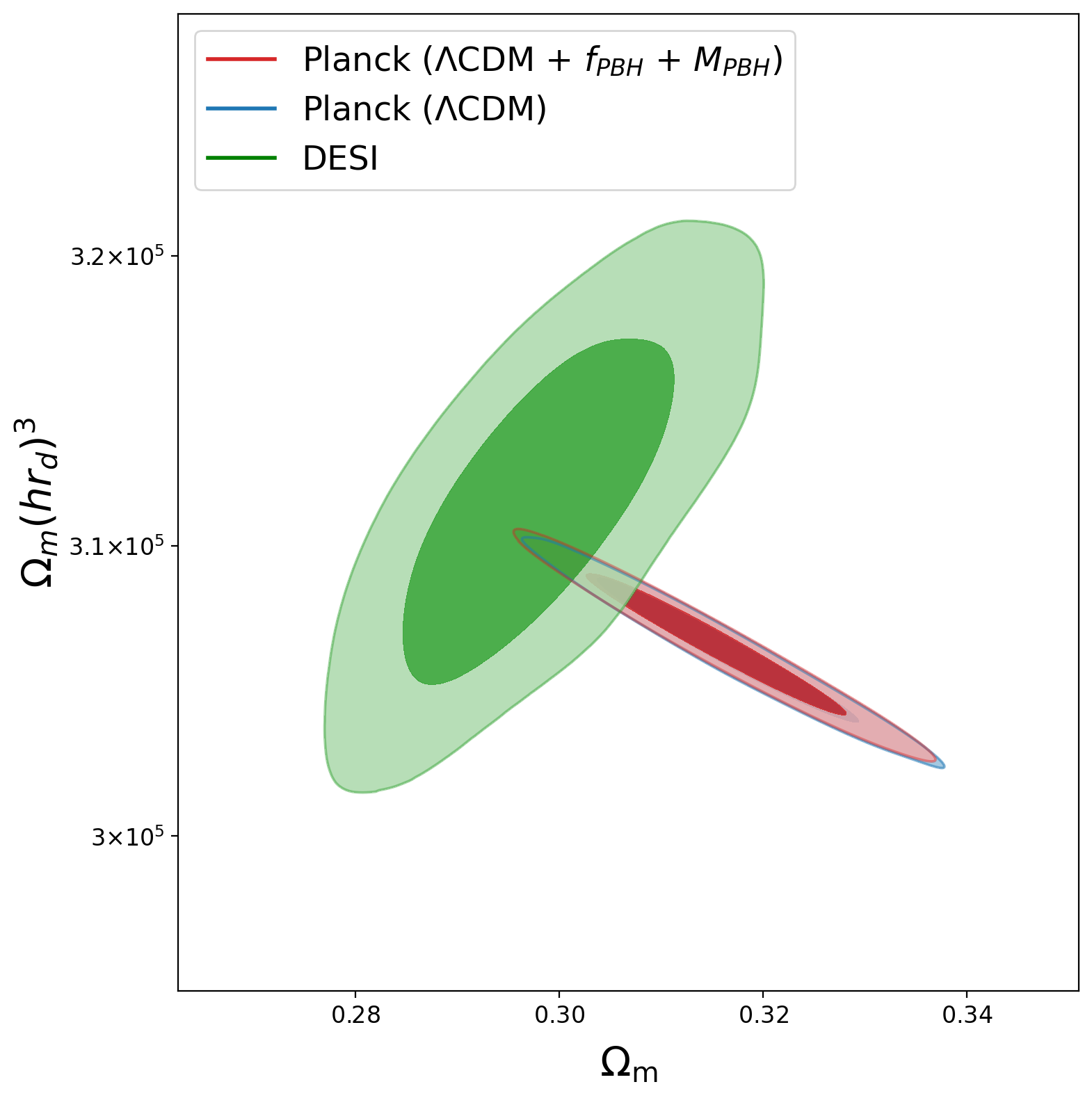}
    \caption{Marginal 2D posterior in the $\Omega_m$ - $\Omega_m (h r_d)^3$ plane, shown  to illustrate the marginal improvement in the BAO--CMB tension.}
    \label{fig:BAOplane}
\end{figure}

\begin{table*}
\centering
\renewcommand{\arraystretch}{1.15}
\begin{tabular}{lc @{\hskip 18pt} c @{\hskip 18pt} c @{\hskip 18pt} c @{\hskip 18pt} ccc}
\hline\hline
Model & $\Delta\chi^2$ & $\tau$ & $\omega_{\rm c}$ & $\Omega_m$ & \mpbh $/10^{14}$g & $10^{11}f_{\rm PBH}$ & $R-1$\\
\hline
$\Lambda$CDM & -- & $0.055^{+0.005}_{-0.008}$ & $0.120_{-0.001}^{+0.001}$ & $0.316^{+0.008}_{-0.008}$ & -- & -- & 0.006 \\
$\Lambda$CDM $+\,f_{\rm PBH}+\mpbh$ & $-0.59$ & $0.063^{+0.007}_{-0.009}$ & $0.120_{-0.001}^{+0.001}$ & $0.315^{+0.008}_{-0.009}$ & $>2.28$ & $< 452$ & 0.001 \\
\hline\hline
\end{tabular}
\caption{Some results given our baseline likelihood alone, Planck. The top row is for \lcdm\ and the next row is our baseline analysis with \lcdm\ extended to include $f_{\rm PBH}$ and \mpbh. The chains are well converged, as quantified by the Gelman-Rubin statistic $R-1$. We see that $\tau$ shifts upward some in the broader model space, while $\Omega_{\rm m}$ and $\omega_{\rm c}$ change almost not at all. The $\Delta \chi^2$ refers to the difference between the best fitting sample in the chain and the corresponding $\Lambda$CDM chain. The improvement in $\chi^2$ by less than 2 indicates no preference for the extension.}
\label{tab:mcmc_models}
\end{table*}

\section{Results}\label{sec:results}

Despite the additional degrees of freedom introduced by our PBH evaporation model, our combined statistical analyses reveal that the standard constraints on the reionization optical depth, $\tau$, remain robust, severely restricting the allowed energy deposition from these PBH sources. Although the $\tau$ posteriors shift somewhat toward larger values, the shifts are less than a third of the desired $\Delta \tau \simeq 0.03$. This is true for both the mean $\tau$ and 95\% confidence upper limits. As expected, there are nearly no shifts in $A_s \exp(-2\tau)$ so the $\Delta \tau$ leads to a small $\Delta A_s/A_s \simeq 2\Delta \tau$. Surprisingly, however, there are almost no shifts in the remaining cosmological parameters. In Section~\ref{sec:Shifts} we discuss these 
shifts in parameters in more detail before commenting on their relevance for the matter density deficit and lensing excess in Section~\ref{sec:MatterAndLensing}. We conclude this section with a discussion of the sensitivity of our results to assumed prior distributions in $f_{\rm PBH}$ and \mpbh.

\subsection{Shifts in \LCDM\ parameters}
\label{sec:Shifts}

We now examine how the PBH signal affects the \lcdm\ parameter constraints inferred from Planck data. The biggest change, in units of the standard deviation of the one-dimensional marginal posterior, is in $\tau$. Here the \LCDM\ result of $\tau = 0.055^{+0.005}_{-0.008}$ changes to $\tau = 0.063^{+0.007}_{-0.009}$ with an increase in the 95\% confidence upper limit from 0.068 to 0.079 \footnote{We found that using \texttt{SRoll2} instead of our baseline \texttt{SimAll} for low-$\ell \ EE$ resulted in a further upward shift in $\tau$ by 0.003. This is similar to the size of shift seen in \lcdm\ when switching between these two likelihoods \citep{Pagano:2019tci}.}.

The next biggest change is to $A_s$, by an amount that keeps $A_s \exp(-2\tau)$ nearly constant. Beyond $A_s$ and $\tau$, the other shifts are negligible. There is a very slight shift of $\omega_c$ (and $\Omega_m$) to lower values. This is the direction we expect due to the increase in $A_s$ and the positive response to lensing power of increases in either $A_s$ or $\omega_c$ (or $\Omega_m$).

In the top right of Fig.~\ref{fig:Cornerplot} we zoom in on the $\Omega_m$ vs. $\tau$ plane where we can more clearly see that the shift in this 2D marginal posterior is almost entirely in $\tau$, with a very slight movement toward the DESI-preferred value of $\Omega_m$. This shift in mean $\Omega_m$ is only 5.7\% of the difference between the Planck \lcdm\ and DESI means.

We show a more complete comparison with DESI in Fig.~\ref{fig:BAOplane}. In \lcdm, BAO data can be losslessly compressed to $\Omega_m$ and $h r_d$. Therefore they can also be losslessly compressed to $\Omega_m$ and $\Omega_m (h r_d)^3$, a parameter pair with a significantly reduced correlation. The reduced correlation makes it easier to see the very small shift, though it remains subtle. The difference between Planck and DESI in this 2D plane shifts from 1.70$\sigma$ to 1.58$\sigma$, with this 2D tension quantified using the Gaussian parameter-difference statistic (the Mahalanobis distance \citep{Mahalanobis1936}), which assumes 2D Gaussian posteriors.\footnote{The reader may wish to recall that, as mentioned above, this 1.70$\sigma$ difference grows to near 3$\sigma$ when the data are extended to SPA plus SPA Lensing \cite{SPT-3G:2025bzu}.}
  
The small shifts in the parameters other than $\tau$ and $A_s$ are surprising. While $\tau$ shifts by about one third of the magnitude required to completely resolve the BAO--CMB tension, the corresponding shifts in other parameters are even smaller by this same metric. As already stated, we do see the upward movement in $A_s$ expected from the upward shift in $\tau$. We expected, as is seen in \lcdm\ when the $\ell < 30$ data are ignored \cite{Sailer:2025lxj,Jhaveri:2025neg}, that this upward shift in $A_s$ would result in a reduction of $\omega_{\rm c}$ (and therefore $\omega_m$ and $\Omega_m$). This expectation is due to the boost in lensing from the increase in $A_s$ which then allows for $\omega_c$ to relax downward, as the peak-smoothing in the Planck data no longer drives it up as strongly. 

Clearly, these expectations do not play out in the actual results. The relaxation of $\omega_{\rm c}$ described above is the \lcdm\ response to a \emph{reionization-like} increase in $\tau$; that is, a nearly uniform suppression of power at $\ell > 30$ together with the generation of large-scale polarization. Our analysis, however, varies $z_{\rm reio}$ and retains the low-$\ell$ $EE$ data, and in that setting, such a signal moves no parameter but $z_{\rm reio}$ itself. Furthermore, while the PBH signal is not exactly ``reionization-like", there is substantial similarity between the two ways of boosting $\tau$. This means that adding a population of PBHs on top of a fixed \lcdm\ model raises the large-scale polarization power beyond what the low-$\ell$ $EE$ data prefer, and in the fit, $z_{\rm reio}$ decreases to compensate. This $z_{\rm reio}$ adjustment compensates for roughly half of the optical depth supplied by PBH evaporation, so the net optical depth of the fitted model rises by the remainder. See Appendix~\ref{sec:fisher_investigation} for further discussion.

Since $z_{\rm reio}$ is free to adjust, shifts in other parameters, particularly $\omega_{\rm c}$, must be driven by the ways in which the PBH signal differs in shape from a shift of the reionization step. Those differences, discussed in Section~\ref{sec:ImpactOnCMB} and illustrated in Fig.~\ref{fig:Spectra}, manifest as structure at multipoles above the reionization bump that no adjustment of $z_{\rm reio}$ reproduces. This structure drives two competing responses in $\omega_{\rm c}$. The lensing-mediated response persists --- $A_s$ still rises in accordance with the net rise of the optical depth, and with it the predicted lensing power --- but there is now an upwards push on $\omega_{\rm c}$ as well. Over the evaporation redshifts (i.e. masses) that the data allow, these two responses nearly cancel, and the net movement of $\omega_{\rm c}$ is negligible. Meanwhile, the small upward shift we do find in $\tau$ reflects the prior volume opened by our one-sided extension rather than a preference of the data (see Section~\ref{sec:Prior}). Regardless, that small increase arrives without the accompanying relaxation downwards of $\omega_{\rm c}$.

\subsection{Matter density deficit and lensing excess}
\label{sec:MatterAndLensing}

\begin{table*}
\centering
\renewcommand{\arraystretch}{1.15}
\resizebox{\textwidth}{!}{
\begin{tabular}{l @{\hskip 10pt} l @{\hskip 18pt} c @{\hskip 18pt} c @{\hskip 18pt} c @{\hskip 18pt} ccccc}
\hline\hline
Model & $\Delta\chi^2$ & $\tau$ & $\omega_{\rm c}$ & $\Omega_{\rm m}$ & \mpbh ($10^{14}$ g) & $10^{11}\,f_{\rm PBH}$ & $A_{\rm lens}$ & $R-1$\\
\hline
\multirow{1}{3.5cm}{$\Lambda$CDM $ + \ A_{\rm lens}$}
 & -- & $0.054^{+0.004}_{-0.008}$ & $0.117^{+0.001}_{-0.001}$ & $0.299^{+0.004}_{-0.004}$ & -- & -- & $1.081^{+0.025}_{-0.025}$ & $0.011$ \\
\hline

\multirow{1}{3.5cm}{$\Lambda$CDM $+\,f_{\rm PBH}+A_{\rm lens}$}
 & $1.50$ & $0.056^{+0.003}_{-0.006}$ & $0.117^{+0.001}_{-0.001}$ & $0.299^{+0.003}_{-0.005}$ & $4.0$ & $< 1348$ & $1.068^{+0.022}_{-0.022}$ & $0.007$ \\
\hline\hline
\end{tabular}
}
\caption{Posterior means and marginalized uncertainties for cosmological and PBH parameters in the $\LCDM$ and 
$\LCDM$+PBH models (varying $A_{\rm lens}$), using the SPA + DESI + SPA lensing likelihood combination. The $\Delta\chi^2$ value is
relative to $\LCDM$, taken from the best fitting sample in the chain. The bias from not using the true best fit is estimated to be $\lesssim 2$, so our conclusions should not be impacted.}
\label{tab:mcmc_models2}
\end{table*}

Having motivated our study with the matter density deficit and CMB lensing excess, we now examine the extent to which the PBH reionization scenario addresses them.

\citet{Lynch:2025ine}, following \citet{Loverde:2024nfi}, distinguish two related aspects of the tension between DESI BAO and CMB data. 
The \emph{matter density deficit} is the preference for a smaller $\omega_m r_d^2$ from the acoustic scale (BAO together with the CMB $\theta_{\rm s}$) than the $\omega_{cb} r_d^2$ preferred by the primary CMB spectra, where $\omega_{cb} \equiv \omega_c + \omega_b$.
The \emph{lensing excess} is the preference, in CMB lensing reconstruction and in the peak smoothing of the lensed $TT$, $TE$, and $EE$ spectra, for more lensing power than is predicted by non-lensing information in the primary CMB spectra, particularly when the latter are combined with DESI BAO. The two are related but distinct: \citet{Lynch:2025ine} show, for instance, that allowing a fraction of the dark matter to decay removes the matter density deficit while leaving the lensing excess intact.

We first consider the matter density deficit. The inference of $\omega_m r_d^2$ from the acoustic scale is unaffected by our extension from \lcdm\ to the PBH model space: changes to the reionization history do not affect the interpretation of BAO data, and we have checked that the $\theta_{\rm s}$ posterior is likewise negligibly changed. We therefore adopt the Acoustic Scale constraint of \citet{Weiner:2026sfm} and compare it to the $\omega_{cb} r_d^2$ posteriors under the \lcdm\ and PBH models, both inferred from our baseline Planck likelihood. The marginalized posteriors for these quantities are shown in Fig.~\ref{fig:MatterDeficit}, where the Acoustic Scale curve is taken from \citet{Weiner:2026sfm}.\footnote{The $\theta_{\rm s}$ constraint used here is from the SPA dataset rather than Planck alone.} The deficit is reduced only marginally, from $1.68\sigma$ to $1.57\sigma$. 

Since \lcdm, by definition, includes a contribution to $\omega_{\rm m}$ from neutrinos of the minimal amount expected assuming the normal hierarchy, $\omega_\nu^{\rm min} =$ 0.058 eV/ 93.14 eV, we also examine the matter density deficit with this contribution included. That is, we compare $\omega_m r_d^2$ from the Acoustic Scale to $(\omega_{\rm cb} + \omega_\nu^{\rm min}) r_d^2$ from CMB data. The matter density deficits in both \lcdm\ and with the extension to PBH increase by about the same amount to 2.39$\sigma$ and 2.27$\sigma$ respectively. 

Having established that PBH evaporation does not substantially alter $\omega_{\rm m}$ and the matter deficit, the next question is: how does it affect the lensing excess? To a good approximation the only lensing-relevant parameter the PBHs change is $A_{\rm s}$, which rises to hold $A_{\rm s} \exp(-2\tau)$ fixed, by about $2\Delta\tau$. Even our largest $\tau$ shift---the $\Delta\tau \simeq 0.008$ of our baseline Planck result---therefore raises the predicted lensing power by only about $1.5\%$, and so is expected to reduce the lensing excess by a correspondingly small amount.

We test this with the datasets most sensitive to the excess: SPT-3G and ACT together with their lensing reconstructions, and DESI (the SPA + SPA Lensing + DESI combination of Section~\ref{sec:data}). These data are particularly informative because the lensing reconstructions constrain the excess directly, and because their high-$\ell$ spectra are where the PBH reionization signal leaves its imprint on cosmological parameter inference. As in  \citet{SPT-3G:2024atg}, $A_{\rm lens}$ scales all model lensing spectra from their physical values---both those entering the lensed $TT$, $TE$, and $EE$ spectra and those compared with the reconstructed lensing bandpowers.\footnote{To reduce the model space here we fix $\mpbh = 4\times10^{14}$g and vary only $f_{\rm PBH}$. This lies above the upper edge of the mass prior adopted elsewhere, however, this is deliberate. Ionization histories with peak evaporation redshift $z_{\rm peak} > z_{\rm reio}$ are strongly disfavored by the data, and for $z_{\rm peak} < z_{\rm reio}$, $X_{\rm e}$ rises smoothly at earlier times and the shape of the history is nearly independent of \mpbh, rendering $f_{\rm PBH}$ and \mpbh\ degenerate. This mass is therefore representative of the configurations the data tolerate best.}

We find $A_{\rm lens} = 1.068^{+0.022}_{-0.022}$ in the PBH model, compared with $1.081^{+0.025}_{-0.025}$ under \lcdm\ for the same data. The shift is in the expected direction---the added optical depth raises $A_{\rm s}$ at fixed $A_{\rm s}e^{-2\tau}$, and with it the predicted lensing power, so that a smaller $A_{\rm lens}$ is required---but it is far too small to matter: the excess stands at $3.2\sigma$ under \lcdm\ and $3.1\sigma$ in the PBH model. Closing the lensing excess entirely would require $\Delta\tau \simeq 0.04$, five times the largest shift we obtain anywhere in this work. The remaining parameters relevant to the tension are unmoved, with $\omega_{\rm c} = 0.117$ and $\Omega_{\rm m} = 0.299$ in both models, and the data show no preference for the extension.\footnote{The best-fitting sample in the PBH chain is marginally worse than that of the \lcdm\ chain, $\Delta\chi^2 = +1.50$. Since the models are nested, the true
difference cannot be positive; here we are using the best-fitting sample from each chain as a proxy for the true minimum, and a difference this close to zero indicates that the additional parameter buys no appreciable improvement in the fit.}

\begin{figure} 
    \centering
    \includegraphics[width=\columnwidth]{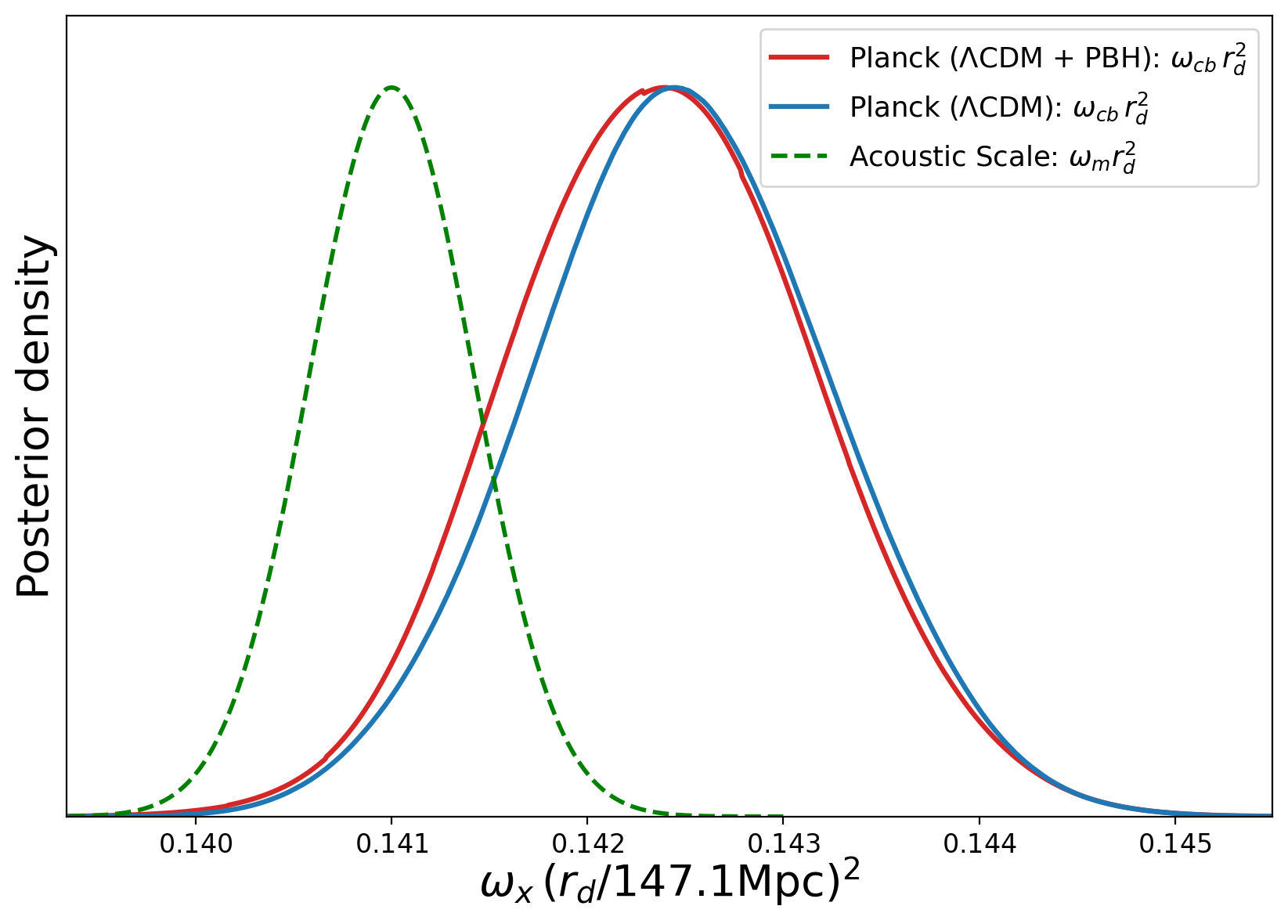}
    \caption{1D posterior for $\omega_x (r_d/147.1 \rm{Mpc})^2$ where $\omega_x = \omega_c + \omega_b$ for the CMB datasets (red and blue solid curves) and $\omega_x = \omega_m$ for the Acoustic Scale dataset (green dashed curve).}
    \label{fig:MatterDeficit}
\end{figure}

\subsection{Prior sensitivity}
\label{sec:Prior}

Here we explore the sensitivity of our posterior distributions to assumed priors on the PBH parameters.

The first question we ask is whether the shift in $\tau$ in the marginalized 1D posterior is due to any preference in the data for higher $\tau$, or is instead an effect of marginalization. The answer appears to be that the data have essentially no preference at all for the PBH signals. The sample in our baseline chain with the highest likelihood has $\tau = 0.054$, very near the \lcdm\ mean of 0.055, and the improvement in fit is $\Delta \chi^2 = -0.59$ for the two additional parameters. This suggests that the shift in the mean of the 1D marginal posterior from $0.055$ to $0.063$ is due instead to the parameter space volume that has been opened up in our one-sided extension, one-sided in the sense that the introduction of PBHs can only increase free electron density, not reduce it. 

This interpretation is supported by our investigation of sensitivity to our priors on $f_{\rm PBH}$. Our baseline results are run with uniform priors on $\mpbh$ and $f_{\rm PBH}$. Switching to a log-uniform prior on $f_{\rm PBH}$ more heavily weights those regions of the parameter space with negligible contribution to $\tau$ and the mean of the marginal posterior reduces to $\tau = 0.058$; the shift above the \lcdm\ mean falls from $0.008$ to $0.003$. Note also that the preference for higher \mpbh\ that we found in the marginal posterior for our baseline case is prior dependent. Switching to a log prior in \fpbh, or a log prior in both \fpbh\ and \mpbh, results in favoring the low mass end of our range over the high mass end. 

That these features move with the prior is itself informative. Because the data express no preference for the PBH signal, the size of the $\tau$ shift is set by the volume of parameter space the prior makes available rather than by the likelihood, and the consequences for the tensions follow directly. 
The uniform priors of our baseline are thus the most favorable case among those we consider, and even there the amelioration is marginal.

\section{Discussion and Conclusions}\label{sec:conclusions}

In this work we explored the possibility of easing tensions between BAO and CMB data that are present in the context of the \lcdm\ model, by extending that model to allow for high-redshift partial reionization. Increasing $\tau$ can lead to an increase in $A_s$, which boosts lensing power predictions, thereby addressing the CMB lensing excess. The boosted lensing power also relaxes downward CMB inferences of $\omega_{\rm c}$, and, critically for addressing the matter density deficit, of $\omega_{\rm cb} r_d^2$. Thus increasing $\tau$ can be a path toward resolving tensions between CMB and BAO data sets \cite{Sailer:2025lxj,Jhaveri:2025neg}. 

For specificity, we took Hawking radiation from a monochromatic spectrum of primordial black holes as the source of the energy injection leading to additional free electrons. We use the code and model of \citet{Poulin:2016anj} with several simplifying choices, some of which are known to impact the resulting ionization histories \citep{Stocker:2018exoclass, Acharya:2020jbv,Yin:2026hvw}. However, the ionization history differences are almost entirely ones of amplitude, rather than shape (see Appendix~\ref{sec:BeyondOn-the-spot}). These assumptions would bias any determination of upper limits on $f_{\rm PBH}$, but we do not expect them to significantly influence our $\tau$ inferences.

We found that, using uniform priors on $f_{\rm PBH}$ and \mpbh, high-redshift reionization from PBHs can shift the 1D marginal posterior for $\tau$, given Planck primary CMB power spectra constraints, upward by about $0.008$---less than a third of the $\Delta\tau \simeq 0.03$ that would restore consistency between BAO and CMB data within \lcdm. This shift is itself largely a consequence of the prior volume opened up by our one-sided extension (PBHs can only add free electrons, never remove them) rather than a preference in the data. Indeed, the maximum-likelihood point remains at essentially the \lcdm\ value of $\tau \simeq 0.054$. The CMB data clearly do not favor these additional signals.

One might expect the matter density deficit and CMB lensing excess to ease by a comparable fraction, but they do not. The matter density deficit eases only marginally, from $1.68\sigma$ to $1.57\sigma$, for our baseline Planck set of likelihoods. To examine the lensing excess, we extended the CMB data to include ACT DR6 and SPT-3G D1 $TT/TE/EE$ spectra and lensing reconstructions from all three of these CMB experiments; the lensing excess reduces somewhat with the $A_{\rm lens}$ parameter shifting toward unity, from $A_{\rm lens} = 1.081^{+0.025}_{-0.025}$  assuming \lcdm\  to  $A_{\rm lens} = 1.068^{+0.022}_{-0.022}$ in the PBH model.  Switching to log priors on the PBH parameters yields even smaller shifts in $\tau$, and correspondingly weaker amelioration of each tension.

Why does even this modest increase in $\tau$ fail to ease the matter density deficit? Two things differ from the \lcdm\ expectation sketched above. First, the PBH electrons are added on top of the usual step-like reionization, so in a joint fit $z_{\rm reio}$ shifts downward to keep the low-$\ell$ bump consistent with the data, returning a substantial fraction of the optical depth the PBHs supply. Per unit of net optical depth realized, the downward push $\omega_{\rm c}$ receives through the lensing channel is much as it would be for a shift in $z_{\rm reio}$; what has changed is the net $\Delta\tau$ itself. Second, because free electrons are deposited at higher redshift, the PBH signal carries structure at multipoles well above the reionization bump (Section~\ref{sec:ImpactOnCMB}) that no shift of $z_{\rm reio}$ reproduces. The $\omega_{\rm c}$ response to this structure is upward, and over the deposition redshifts the data permit, it very nearly cancels the lensing-mediated downward response, leaving negligible net movement in $\omega_{\rm c}$ (Appendix~\ref{sec:fisher_investigation}). At higher deposition redshifts the upward response dominates, but the same structure disfavors those histories, a constraint carried predominantly by the high-$\ell$ polarization data.

Though our analysis was done within the context of the PBH model, some general features may be relevant to other proposed high-redshift ionization sources. Recent work on ``reionization flash" scenarios from Pop~III.1 stars has shown that such ionization histories are consistent with Lyman-$\alpha$ forest and patchy kSZ data, and might be able to evade the low-$\ell \ EE$ bounds, all while delivering $\Delta \tau \sim 0.04$ \citep{Tan:2025cua, Tan:2025obi, Aggarwal:2026ogm}. Consistency with these data is a non-trivial test which the Pop~III.1 model seems to pass, but our work indicates that other data are relevant as well. This is because more complex ionization histories than the standard ``step-like" reionization introduce signals at smaller scales, where high-$\ell$ $TT/TE/EE$ data from Planck, SPT, and ACT are informative. 

For the PBH reionization histories considered here, those signals constrain the model and shift $\omega_{\rm c}$, impacting whether the BAO--CMB tension is truly reduced along with the increased $\tau$. While the corresponding signal from a Pop.~III.1 flash will differ (particularly because the flash is largely confined in redshift between $z \sim 10-30 $), there is \textit{a priori} no reason to expect it to vanish. Additionally, when the reionization redshift is estimated jointly rather than held fixed, it adjusts in response to the added signal component that resembles ordinary reionization, to keep the low-$\ell \ EE$ bump fixed. This reduces the net increase in $\tau$ realized by the fit.

We take these two facts --- that modifications to $X_{\rm e}(z)$ at low redshift are partially degenerate with changes to $z_{\rm reio}$, and that high redshift modifications introduce new signals that must fit the data --- as discouraging of high-redshift reionization as a route to relaxing the BAO--CMB tension through an increased $\tau$. However, our analysis was limited to an essentially one-parameter family of PBH-specific ionization histories, and it remains possible that there are potentially interesting surprises to be found with a more flexible, phenomenological exploration. As we have seen, the specifics of the considered ionization history affect parameter inference: it is possible that in a different model space they could even do so constructively.

Given these discouraging findings, it is worth remarking on other solutions that could boost $\tau$. One class of solutions accommodates a large $\tau$ by lowering or obscuring the observed large-scale polarization power without affecting the ionization history. One example is a suppression of large-scale primordial power, as arises in models with a departure from slow-roll inflation, which lowers the baseline large-scale polarization signal and enables a compensatory enhancement in $\tau$ in order to fit the existing data \citep{Jhaveri:2025neg, Jhaveri:2026qzi}. As pointed out by \citet{Namikawa:2025doa}, another involves cosmic birefringence: a large rotation angle during reionization causes polarization generated at different epochs to acquire different phases, which partially cancel in the line-of-sight solution and suppress the low-$\ell$ EE power. Finally, it remains a possibility that the low-$\ell$ EE uncertainties are underestimated. Reanalyses of large-scale polarization data from Planck and WMAP, which differ in map-making techniques, systematics treatments, and likelihoods, have found central values spanning $\tau \simeq 0.051 - 0.063$ \citep{Planck:2016kqe, Delouis:2019bub, Pagano:2019tci, Genesini:2026lmg, Tristram:2023haj}. We note that none of these values are sufficient to fully resolve the BAO--CMB tension.

Of course, the moderate BAO--CMB tension may have nothing to do with $\tau$. It might be due to pre-recombination changes that reduce $\omega_{\rm cb} r_d^2$, such as early recombination (perhaps due to primordial magnetic fields) \citep{Lynch:2024hzh, Mirpoorian:2025rfp}, or early dark energy \citep{Chaussidon:2025npr,Poulin:2025nfb,SPT-3G:2025vyw,Jhaveri:2026bla}, which decreases $r_d^2$ more than it boosts $\omega_{\rm cb}$, or late-time changes such as evolving dark energy or dark matter - dark energy interactions \citep{2025JZhai_IDE,2024Giare_IDE, Khoury:2025txd}. It could even conceivably be a statistical fluke; the statistical significance is not overwhelming.

Given the intrinsic difficulty of measuring the low-$\ell$ EE polarization signal, we eagerly anticipate cross-checks with Planck from future Cosmology Large Angular Scale Surveyor \citep{2014CLASS_TOM} results in the near future, as well as future cosmic-variance-limited observations from \texttt{LiteBIRD} \citep{2023Litebird} and other proposed space-based CMB instruments. With CMB lensing playing a central role in the BAO--CMB tension, we also look forward to improved lensing constraints very soon from SPT-3G D1 (Omori et al. \textit{in preparation}), to be followed by results from the wide-area SPT-3G Ext-10K survey, and also from Simons Observatory \citep{2019Simons_forecast}; see \citet{SPT-3G:2024qkd} for detailed forecasts of these upcoming lensing measurements. 

 \acknowledgments

 We thank J. Chluba, S. Raghunathan, N. Sailer, M. Gerbino, and M. Kaplinghat for useful conversations. This work was supported in part by DOE Office of Science award DESC0009999 and the Michael and Ester Vaida Endowed Chair in Cosmology and Astrophysics. This project has received funding from the European Research Council (ERC) under the European Union’s Horizon 2020 research and innovation programme (grant agreement No 101001897).
 
\appendix

\section{Linear response analysis of \\ the $\omega_{\rm c}$ shift}
\label{sec:fisher_investigation}

In Section~\ref{sec:Shifts}, we found less of a shift in $\omega_{\rm c}$ towards lower values than the PBH-induced change in $\tau$ and the lensing-mediated $\tau$--$A_{s}$--$\omega_{\rm c}$ degeneracy would suggest. In this appendix, we trace that outcome to the structure of the PBH signal.

\subsection{Framework}

We consider the response of cosmological parameters when fitting a signal added to some fiducial spectrum. We let $\theta$ denote the cosmological parameters varied in the fit, and $\lambda$ the parameters of the added signal, which are held fixed. We then define an inner product on power-spectrum perturbations $A,B$ (e.g. $\delta C_\ell$) from a likelihood
covariance $\mathsf{C}$,
\begin{equation}
\langle A, B\rangle \equiv (\mathcal{B}A)^{\mathsf T}\, \mathsf{C}^{-1}\,(\mathcal{B}B),
\end{equation}
where $\mathcal{B}$ is the bandpower binning operator, so that $\langle X,X\rangle$ is the $\Delta\chi^2$ an unmodeled signal template $X$ would contribute.

Given derivative templates $D_p = \partial C_\ell/\partial \theta_p$ and Fisher matrix $F_{pq} = \langle D_p, D_q\rangle$ (plus priors), a signal $T(\lambda)$ added to the fiducial model displaces the best fit over $\theta$ by
\begin{equation}
\delta\theta(\lambda) = -F^{-1}v, \qquad v_p = \langle T(\lambda), D_p\rangle .
\label{eq:shift}
\end{equation}
This is just the least-squares solution obtained by building the closest approximation to $T$ from the span of the fitted templates. We note that the $\{ D_p \}$ are not mutually orthogonal. Equivalently, each coefficient may be written as a projection onto that parameter's template after the others have been removed from it,
\begin{equation}
\delta \theta_p = -\frac{\langle T, \hat D_p\rangle}{\langle \hat D_p, \hat D_p\rangle},
\qquad \hat D_p \equiv \big(1 - P_{\rm rest}\big) D_p ,
\label{eq:orthog}
\end{equation}
with $P_{\rm rest}$ projecting onto the remaining fitted templates.

We note that while the data contain no PBH signal, the posteriors discussed in Section~\ref{sec:results} explore non-zero $\lambda$, and at each such $\lambda$ the best fit over $\theta$ is displaced by $\delta \theta(\lambda)$ from the fiducial. Averaging over $\lambda$ gives

\begin{equation}
    \Delta\langle\theta_a\rangle \simeq \int d\lambda\; \mathcal{P}(\lambda)\,\delta\theta_a(\lambda),
\end{equation}
where $\mathcal{P}(\lambda)$ is the marginal posterior of the PBH parameters (cf. Fig. ~\ref{fig:Cornerplot}). To see a shift in the marginal mean, the displacement must be non-zero and there must be posterior mass at $\lambda \neq 0$. The prior sensitivity of Section~\ref{sec:Prior} enters through $\mathcal{P}(\lambda)$.

Our base parameter set is $\theta = \{\omega_{\rm b}, \omega_{\rm c}, H_0, \ln( 10^{10}A_{\rm s}), n_{\rm s}\}$ plus the calibration parameter $A_{\rm planck}$ ($\sigma = 0.0025$ prior). We use two metrics: \texttt{plik\_lite} alone ($30 \le \ell \le 2508$, $TT/TE/EE$), and the same with the compressed low-$\ell$ likelihood of \citet{Prince:2021fdv} added.\footnote{These consist of two log-normal bins in $TT$ and three in $EE$. We calibrate the $EE$ block so that $\sigma(\tau) = 0.0065$ matches our \lcdm\ Planck chain.} We work with $\tau_{\rm exc}$, the optical depth in excess of the recombination-only history, which in \lcdm\ reduces to $\tau_{\rm reio}$.

\subsection{Signal templates}

We compare two routes to the same excess optical depth, $\Delta\tau_{\rm exc}$. The first moves the tanh step in the CAMB parameterization,
\begin{equation}
T_{\rm tanh} = \frac{\partial C_\ell}{\partial \tau_{\rm exc}}\bigg|_{f_{\rm PBH}=0}
\times \Delta\tau_{\rm exc},
\end{equation}
evaluated by central difference in $z_{\rm reio}$ about the fiducial model. The second holds $z_{\rm reio}$ fixed and adds a PBH population with peak evaporation at redshift $z_{\rm peak}$,
\begin{align}
T_{\rm PBH}(\lambda) &= \frac{\partial C_\ell}{\partial \tau_{\rm exc}}\bigg|_{z_{\rm reio}, z_{\rm peak}}
\times \Delta\tau_{\rm exc}, \\ \lambda &\equiv(z_{\rm peak}, \Delta \tau_{\rm exc}),
\end{align}
evaluated with a second-order one-sided stencil, since PBHs can only add electrons. The parameter set $\lambda$ is a reparameterization of $(f_{\rm PBH}, \mpbh)$ into an amplitude and a shape parameter, and all responses quoted below are for $\Delta \tau_{\rm exc} = 0.01$.

\subsection{The lensing-mediated relaxation}

We first reproduce the expected behavior from \citet{Sailer:2025lxj} and \citet{Jhaveri:2025neg}, where we drop the $\ell<30$ data and the step-like reionization $z_{\rm reio}$ is held fixed. We find $\delta\omega_{\rm c}(T_{\rm tanh}) = -2.95\times10^{-4}$ per $\Delta\tau_{\rm exc}=0.01$: raising $\tau$ raises $A_{\rm s}$ at fixed $A_{\rm s}e^{-2\tau}$, raising the lensing power and allowing $\omega_{\rm c}$ to relax. We verify that this response is almost entirely lensing-mediated: adding the lensing amplitude $A_{\rm lens}$ to the set of fitted parameters collapses the response by a factor of $\sim14$, to $-0.218\times10^{-4}$.

\subsection{PBH-induced shifts in the matter density}

\begin{figure}[t]
    \centering
    \includegraphics{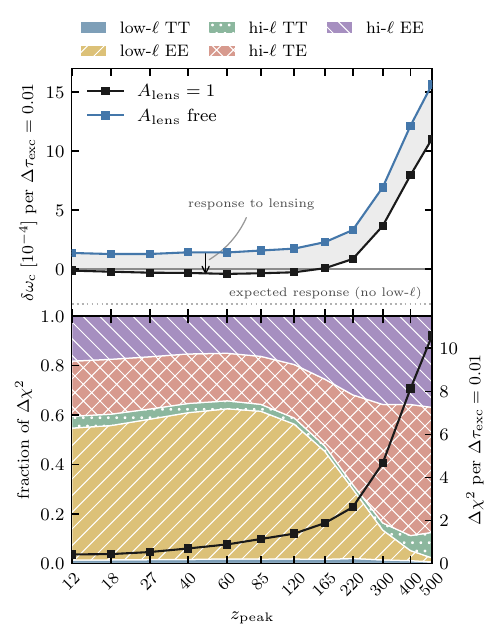}
    \caption{\textbf{Top}: the parameter response of $\delta\omega_{\rm c}$ (in units of $10^{-4}$) to an injected PBH signal with $\Delta \tau_{\rm exc} = 0.01$, computed using a linear-response fit, with the lensing amplitude held at $A_{\rm lens}=1$ (black curve) and with $A_{\rm lens}$ floating freely (blue curve). The shift from the blue to black curve represents $\omega_{\rm c}$ responding to the increased lensing power that results from adjusting $A_s$ in the fit. The dotted line shows the expected relaxation for the different configuration in which $z_{\rm reio}$ is fixed and the low-$\ell$ data are dropped. \textbf{Bottom}: decomposition of the total $\Delta\chi^2$ (per $\Delta\tau_{\rm exc}=0.01$; black curve, right axis) contributed by the signal in the $A_{\rm lens}=1$ fit, split into the fractional contribution from different data blocks.}
    \label{fig:omegac_shifts}
\end{figure}

We now add the compressed low-$\ell$ block to the metric and $z_{\rm reio}$ to the fitted set, and consider the response to $T_{\rm PBH}$. The low-$\ell$ $EE$ data pin the amplitude of the reionization bump, and $z_{\rm reio}$ can be adjusted to maintain that part of the fit. This configuration most closely approximates our main analysis in Section~\ref{sec:results}, where $z_{\rm reio}$ is a sampled parameter and the low-$\ell$ data are used in the fit.

Before considering the response to $T_{\rm PBH}$, we note that in this configuration $T_{\rm tanh}$ is nulled: the only response to this template is from $z_{\rm reio}$. Since $T_{\rm tanh} \propto D_{z_{\rm reio}}$ by construction, the fit can cancel it with a shift $\delta z_{\rm reio}$, leaving zero residual. Any parameter response (other than $\delta z_{\rm reio}$) to a PBH signal that delivers the same $\Delta \tau_{\rm exc}$ is therefore a consequence of the PBH signal's differing shape.

Because the PBH signal's spectral shape differs from the tanh template, it does not project purely onto $D_{z_{\rm reio}}$: for $z_{\rm peak}=12$, the $z_{\rm reio}$ response translates to a shift in $\tau_{\rm reio}$ by only $5.2\times10^{-3}$, roughly half of the $\Delta\tau_{\rm exc}=0.01$ supplied by the injected PBH template, and the optical depth of the fitted model rises by about the same amount. Additionally, in this case other cosmological parameters also respond due to the non-zero overlap $\langle D_p, T_{\rm PBH} \rangle$; we discuss the $\omega_{\rm c}$ response below. Some portion of the signal is orthogonal to the span of ${D_p}$, so the fit instead carries it as a residual balanced across low and high $\ell$ — it is this orthogonal residual that constrains the PBH scenario. The lower panel of Fig.~\ref{fig:omegac_shifts} shows the $\Delta \chi^2$ that this residual costs and how it is distributed across different subsets of the data. It grows steeply with $z_{\rm peak}$, bounding $f_{\rm PBH}$ and disfavoring high-redshift evaporations.

The response of $\omega_{\rm c}$ in particular is small: $\delta\omega_{\rm c}(T_{\rm PBH}) = -0.15\times10^{-4}$ at $z_{\rm peak}=12$, in agreement with the chains. This is not, however, because it has no effect on lensing. Adding $A_{\rm lens}$ to the fitted parameters instead gives $\delta\omega_{\rm c}(T_{\rm PBH}) = +1.35\times10^{-4}$: with $A_{\rm lens}$ free to take up the lensing-like part of the signal, the response of $\omega_{\rm c}$ becomes positively correlated with the PBH signal. This is the structural response of $\omega_{\rm c}$ to the PBH signal beyond lensing that remains. The difference between the two fits, $-1.50\times10^{-4}$, is therefore the size of the lensing-mediated push $\omega_{\rm c}$ absorbs when $A_{\rm lens}$ is held at unity, and it is about half the lensing-mediated relaxation found above with $z_{\rm reio}$ held fixed. Both fits are shown against $z_{\rm peak}$ in the top panel of Fig.~\ref{fig:omegac_shifts}. As $z_{\rm peak}$ grows, this positive-response remainder grows much faster than the lensing-mediated push does, so above $z_{\rm peak}\simeq160$ the two no longer cancel and the net response turns positive.

\begin{figure*}
    \centering
    \includegraphics[scale=0.4]{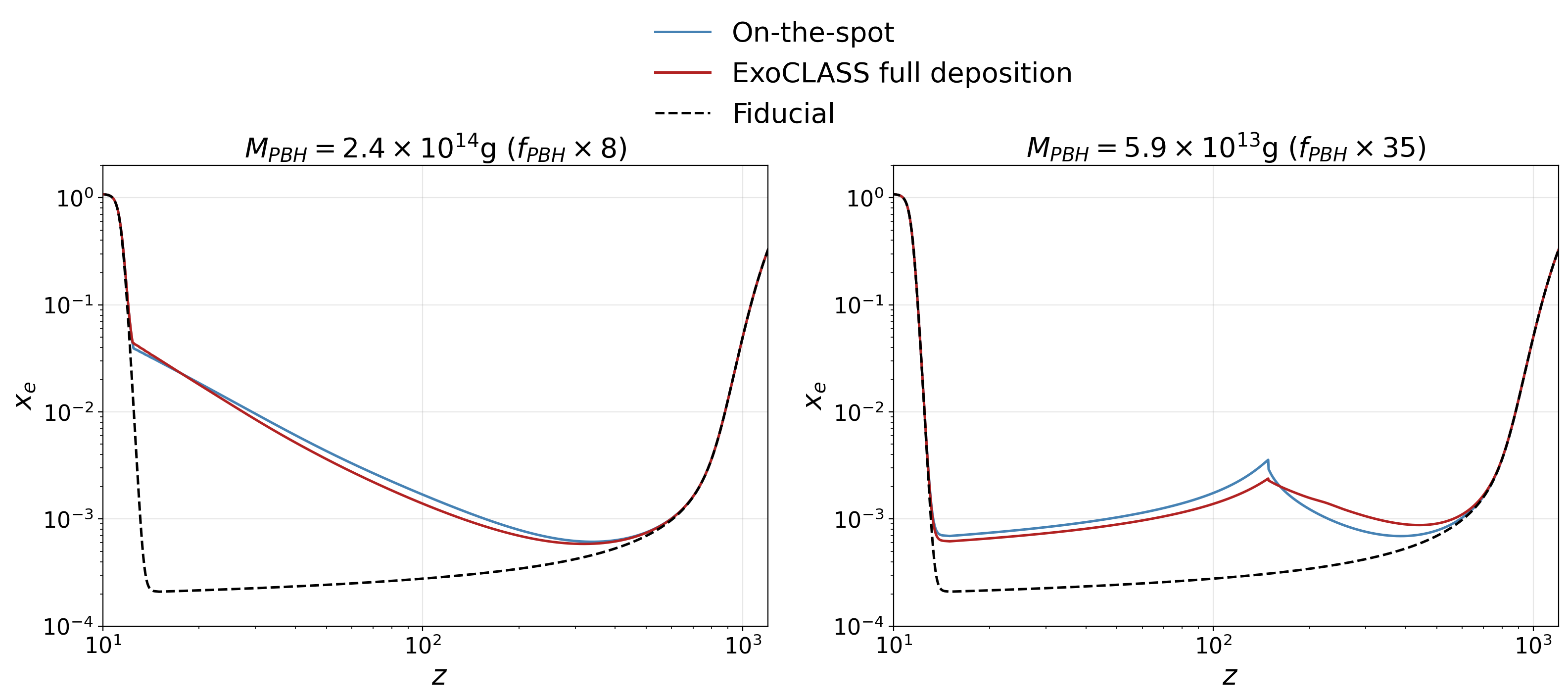}
    \caption{The best-fit ionization histories for the high and low mass models compared to the ExoCLASS output for the same fit, with just the $f_{\rm PBH}$ parameter scaled up in the ExoCLASS scenario in both cases, illustrating the near-degeneracy between the two parameterizations.}
    \label{fig:deposition_approximations}
\end{figure*}

\section{Beyond on-the-spot treatment}
\label{sec:BeyondOn-the-spot}

To validate our instantaneous energy injection approximation, we compared our model against a full electromagnetic cascade treatment using \texttt{ExoCLASS}. This complete treatment relaxes the on-the-spot approximation by accounting for cosmic expansion, particle mean free paths, and redshift-dependent energy deposition. Despite the nuances of this more complicated treatment, we find that for the specific parameter space probed in this work, the morphological differences in the resulting ionization history, $X_{\rm e}(z)$, are minimal. We show examples of ionizations computed under each approximation, for high and low evaporation redshifts, in Fig.~\ref{fig:deposition_approximations}.

The full cascade primarily manifests as a global suppression of the reionization impact, driven by energy channeling into non-ionizing modes, rather than significant structural shifts in $X_{\rm e}(z)$. Because the overall shape of the ionization history remains remarkably consistent between the two treatments, the resulting effects on the CMB visibility function and power spectra are virtually indistinguishable. Consequently, the bounds on the evaporation fraction $f_{\rm PBH}$ derived from the full cascade exhibit a strong effective degeneracy with our simplified model space, with the only physical consequence being slightly tighter constraints on $f_{\rm PBH}$ in our results than with the full deposition treatment.

\bibliography{main}
\end{document}